\documentstyle[12pt]{article}

\topmargin
-2cm
\textwidth
15cm
\textheight
25cm
\oddsidemargin
1cm
\evensidemargin
1cm
\begin{document}
\newcommand{\newc}{\newcommand}
\newc{\sm}{Standard
Model}
\date{}
\title{ Strings, unification
and dilaton/moduli- induced
SUSY-breaking}
  \author{ L.E.
Ib\'a\~nez  \\ \\  Departamento de F\'{\i}sica
Te\'orica, \\  Universidad Aut\'onoma
de Madrid, \\   Cantoblanco,
28049 Madrid, Spain. \\
}
\maketitle
\vspace{-3.5in}
\hspace{4in} FTUAM95/15 ; hep-th/9505098
\vspace{4in}
\begin{abstract}

 I discuss several issues concerning the use of string models as
 unified theories of all interactions.
After a short review of
gauge coupling unification in the string context, I discuss possible
motivations for the construction of  $SU(5)$
and $SO(10)$ String-GUTs. I describe the construction of
such  String-GUTs using different orbifold techniques
and emphasize those properties which could be general.
Although $SO(10)$ and $SU(5)$ String-GUTs are relatively easy to build,
the spectrum bellow the GUT scale is in general bigger than that of the
MSSM and includes colour octets and $SU(2)$ triplets. The phenomenological
prospects of these theories are discussed. I then turn to discuss
 soft SUSY-breaking terms obtained under the assumption of dilaton/moduli
dominance in SUSY-breaking string schemes.  I underline the unique finiteness
properties
of the  soft terms induced by the dilaton sector.
These improved finiteness properties seem to be related to the underlying
$SU(1,1)$ structure of the dilaton couplings.
I conclude with an outlook and
some speculations regarding $N=1$ duality.

\end{abstract}
\maketitle

\bigskip

\bigskip

\centerline{\it Talk at Strings 95, USC, March 1995}


\newpage

\section{Introduction}

Heterotic 4-D strings are considered today our best candidates for
the construction of unified theories of all interactions including
gravity. In spite of that, most of the effort in string theory has been devoted
to understand the theory itself rather than to explore whether indeed one
can really unify the standard model and gravity into a unique consistent
framework. I discuss in this talk some of the ideas considered recently
in this direction, using strings as unified theories.  This is not
supossed to be a general review and I will only
discuss topics in which I have been more or less involved. These include
gauge coupling unification, string-GUTs and  soft SUSY-breaking terms
from dilaton/moduli-induced SUSY-breaking. The las two sections
include some discussion about the special properties of
the the soft terms implied by dilaton-induced SUSY-breaking
and an outlook including some speculations.

\section{Some thoughts about gauge coupling unification}

In 4-D heterotic strings the strength of both gauge and gravitational coupling
constants is goberned by the vev of the dilaton, $<ReS>=4\pi /g^2$.
Consider for example an hypothetical 4-D string whose gauge group contains
the SM group. The corresponding SM couplings would be related as \cite{gins}
\begin{equation}
g_3^2k_3\ =\ g_2^2k_2\ =\ g_1^2k_1\ =\ G_{Newton}M_{String}^2\ =\ {{4\pi}\over
{ReS}}\ .
\label{coup}
\end{equation}
at the string scale $M_s\simeq 5.3g10^{17}GeV$ \cite{kaplu} .
Here $k_2,k_3$ are positive integers (the "levels" of the $SU(2)$ and $SU(3)$
algebras)
and $k_1$ is a rational number which gives the normalization of the $U(1)_Y$
weak hypercharge. If the corresponding 4-D string was constructed by a simple
compactification from either of the two supersymmetric 10-D heterotic
strings, one always has $k_2=k_3=1$ whereas $k_1$ is a model-dependent
rational number. In this way one has at the string scale $sin^2\theta
_W=1/(1+k_1)$ and $\alpha_s = (1+k_1)\alpha _e$, where $\alpha _{s,e}$
are the strong and  electromagnetic fine-structure constants. The
standard $SU(5)$ GUT predictions are recovered for the choice $k_1=5/3$.
As a first try one can assume this value for $k_1$ and compute
$sin^2\theta _W$ and $\alpha _s$ at the weak scale $M_Z$ by running the
renormalization group equations (r.g.e.) down to low energies. Assuming
that the only massless particles charged under the SM are those of the
minimal supersymmetric SM (MSSM) one finds \cite{ilr,anton}
$sin^2\theta _W=0.218$ and $\alpha _s=0.20$,
several standard deviations away from the experimental values
$sin^2\theta _W(M_Z)=0.233\pm 0.003$ and $\alpha _s(M_Z)=0.11 \pm 0.01$.
Is this a serious problem for the idea of a direct string unification
of the SUSY standard model?

One may argue that, for an $SU(3)\times SU(2)\times U(1)$ string
the value of $k_1$ should be considered as a free parameter \cite{prev,iki} .
 If we
indeed take  $k_1$ as a free parameter one can find much more successfull
results for the coupling constants. In particular,for $k_1=1.4$ one
gets  $sin^2\theta _W(M_Z)= 0.235$ and $\alpha _s(M_Z)=0.13$,   in not
unreasonable agreement  with experiment \cite{iki} .
In fact it is amussing to note that, if the historical order of theoretical
ideas
would have been slightly different, the joining of coupling constants could
have
been
considered an outstanding success of string theory! Indeed, imagine the
stringers of the
early seventies would have discovered the supersymmetric heterotic strings
before GUTs (and
SUSY-GUTs) would have been introduced. Georgi, Quinn and Weinberg would have
told them
how to extrapolate the couplings down to low energies. However they would not
have had
any prejudice concerning the value of $k_1$ and they would have taken it as a
free
parameter. Then they would have found that for $k_1=1.4$ one gets
the above successfull results. This could had been
then interpreted as a great succsess of  string theory! (In fact even the
results stated
above for $k_1=5/3$ would have been considered rather successful given the
experimental
precission in the late seventies).

Let us forget now about virtual history and come back to the question whether
direct string unification of the supersymmetric SM  is or not a problem.
Leaving
$k_1$
as a free parameter is certainly a possibility. It is true however that up to
now nobody has constructed any string model with $k_1/k_2< 5/3$, which is what
it seems
to be required. For example, in the $k=1$ orbifold models constructed up to now
one has $k_1 \geq 5/3$. $E_6$-like models ( (2,2) $E_8\times E_8$
compactifications)
have $k_1=5/3$ and often bigger results are obtained for (0,2) orbifold models.
In fact, straightforward compactifications (e.g., level=1) of the heterotic
strings
are likely to yield always $k_1\geq 5/3$
\cite{difar} , but this is not necesarily the case
in all generality. In particular, in higher KM level models one can think
of a value $k_1/k_2\leq 5/3$ even though $k_1>5/3$.
It would be very interesting to find model-independent constraints on $k_1$ or
else find   examples with $k_1/k_2\leq 5/3$.

There are of course, other alternatives  to understand the disagreement found
for the joining of coupling constants. The infinite massive string states
can give rise to substantial one-loop corrections
\cite{kaplu,thres}
to the gauge coupling constants
(string threshold corrections). Indeed this possibility has been studied and
the potential exists for such corrections to explain the discrepancy
\cite{ilr,il}
. Each coupling
gets one-loop corrections as
\begin{equation}
{{4\pi } \over {g_i^2(M_s)}}\ =\ k_iReS \ +\ \Delta ^i_{Th} \ \ \ , i=1,2,3
\label{thres}
\end{equation}
where $ReS=4\pi /g^2$ and $\Delta ^i_{Th}$ are the threshold corrections.
There are two type of threshold corrections: field-independent and
field-dependent. The first of these are expected to be small
and indeed this smallness has been confirmed in explicit computations
for some $(0,2)$ orbifold examples
\cite{kaplu,odos} . The field-dependent threshold corrections
may on the contrary be large depending on the values of the fields. The fields
relevant in
this case are those related to marginal deformations of the underlying CFT like
the moduli $T_a$, $U_a$ fields. These fields parametrize the size (R) and shape
of the
six-dimensional compactification variety (orbifold, Calabi-Yau manifold).
The corresponding threshold functions $\Delta ^i_{Th}(T_a,T_a^*,U_a, U_a^*)$
have been computed in a variety of 4-D strings
\cite{kaplu,thres,odos,KLDOS}
. It seems to be a common feature
that for large compactification radius $R^2$ one gets for all the examples
studied
a leading correction of the form \cite{IN}
\begin{equation}
\Delta ^i\ =\ {{b'_i}\over {12}}\ R^2
\label{ibnil}
\end{equation}
where the $b_i'$ are model-dependent constant coefficients
\cite{thres}
. Now one can see that
for $k_1=5/3$ and not too large values of $R$ (e.g. $R\simeq 2-4$) one can
obtain good results for $sin^2\theta _W$ and $\alpha _s$ by apropriately
chosing the $b_i'$ coefficients
\cite{ilr,il}
. In fact this result is less trivial than it
sounds since these $b_i'$ coefficients are numbers which may be computed in
specific (orbifold) models in terms
of the quantum numbers and "modular weights" of the  massless field of the
theories.
It was found in particular that only a quite limited class of orbifold models
\cite{il}
could possibly have adequate threshold corrections. A random distribution of
modular weights for the SM particles will in general lead to threshold
corrections
in the wrong direction. So even though indeed threshold corrections may be
large
there is no reason for them to conspire precisely in the way we want.

Another obvious alternative is to give up the idea of a direct string
unification
of the SM involving just the content of the MSSM, i.e., to consider the
presence of extra massless particles in addition to those of the SM
\cite{anton,ram}
. After all,
the explicit 4-D string models constructed up to now have always plenty of
additional stuff! The problem with this is that we open a Pandora box of
virtually unlimited possibilities in which we, rather than predicting the weak
angle,
are just adjusting it. A second problem with this is that explicit computations
of the running couplings for different possible choices of extra matter fields
have shown that there is no posible choice of extra matter fields which yields
direct unification at the string scale {\it  new intermediate scale
thresholds are necessary}
\cite{anton,ram,difar}
. Thus we lose the beauty of direct string unification
altogether. But, if we have to deal with extra intermediate scales, why not
considering the possibility of GUTs themselves which naturally require a scale
of order $10^{16}$ GeV?

\section{Are there motivations for constructing GUTs from strings?}

Standard SUSY-GUTs like $SU(5)$ and $SO(10)$ predict the unification of
coupling
constants at a scale $M_X$ which is a free parameter. This allows for the
computation of
one coupling constant as a function of the others. One finds $\alpha
_s(M_Z)=0.12$ for
$sin^2\theta _W(M_Z)=0.233$ by chosing $M_X=2\times 10^{16}$ GeV, in good
agreement with experiment. Given this success, in principle
much better than the one obtained from direct string unification of the MSSM,
it is natural to try and embed standard SUSY-GUTs into string theory.
This is obviously an important motivation. However, I must remark that in order
for a SUSY-GUT to yield the above nice prediction for $\alpha _s$, it is
crucial
the assumption that bellow the string scale the only particles present in the
spectrum are those of the MSSM. So  unification into a simple group like
$SU(5)$ or $SO(10)$ is not enough, the breaking of those groups has to be such
that
the remaining low energy theory is the MSSM.

One could also consider as an argument in favour of GUTs the nice way in which
the observed SM generations fit into representations of the $SU(5)$ or $SO(10)$
groups. A random 4-D string with $SU(3)\times SU(2)\times U(1)\times G$ group
typically contains extra vectorlike  heavy leptons or quarks. These extra
particles
are in fact chiral with respect to the extended group $G$ and remain light
as long as the extended symmetries are unbroken. Furthermore, these extra
leptons
and quarks are not guaranteed to have integer charges or to obey the usual
charge quantization of the SM particles. In order to obtain (level=1)
$SU(3)\times SU(2)\times U(1)\times G$ string models in which the light
fermions are just ordinary SM generations with standard charge asignements
the simplest way is to start with heterotic strings with some underlying
(level=1) $SU(5)$, $SO(10)$ or $E_6$  symmetry which is broken down to
some smaller group {\it at the string scale} through the Hosotani-Witten
flux-breaking
mechanism. Notice that these models are {\it not} GUTs, because the symmetry
breaking is not
carried out through a Higgs mechanism (i.e., through an adjoint vev) and also
because the GUT-like symmetries are never realized as GUT symmetries at any
scale,
they constitute just an intermediate step in the construction technique.
The actual gauge symmetry is the one of the SM (or some simple extension) and
hence
they are just SM strings of a particular class. Anyway, if we need to have at
some
level some GUT-like structure (although not realized as a complete gauge
symmetry)
it is reasonable to try and study whether symmetries like $SU(5)$ and $SO(10)$
may be promoted to a complete symmetry of the massive spectrum. This would be
the essence of string GUTs.

There are other features of SUSY-GUTs which their practitioners love, like
the prediction for the $m_b/m_{\tau}$ ratio and other fermion mass
relationships;
predictions for proton-decay and lepton-number violating processes etc. These
are
more model dependent and may also be present in string models so they would not
constitute by themselves motivations to construct string GUTs. On the other
hand
it would be usefull to construct GUTs from strings to check whether the
dynamical assumptions that the GUT practitioners assume are or not natural
within the context of strings. So one can try to extract some selection rules
to
constrain
the rules of SUSY-GUT model-building. I will briefly discuss some of these in
the next
section.

\section{SUSY-GUTs from strings}

It is essential for SUSY-GUTs the existence in the spectrum of chiral fields
(e.g., adjoints)
appropriate to induce the breaking of the gauge symmetry down to the standard
model.
In the context of N=1, 4-D strings  this is only possible if the affine Lie
algebra associated to the GUT symmetry is realized at level $k\geq 2$.
Straightforward
compactifications of the supersymmetric heterotic strings have
always $k=1$ algebras  which they inherit from the
$k=1$ $E_8\times E_8$ or $Spin(32)$ $D=10$ heterotic strings.
To obtain 4-D strings with higher level one has to go beyond simple
compactifications
of the heteorotic strings. At the begining it was thought that such higher
level
models would be very complicated to construct.
This is why in the early days of string model-building there were no attempts
at
the
construction of string-GUTs. Only a few papers dealt with the explicit
construction of
4-D strings with affine Lie algebras at higher levels \cite{lew,fiq} .

In the last year there have been
renewed attempts for the construction of string GUTs at $k=2$ using orbifold
\cite{nos}
 and free
fermion techniques
\cite{others} . The first of these methods is relatively easy  and is the one
I am more familiar with. Furthermore world-sheet supersymmetry (which is quite
a
technical difficulty in fermionic models
\cite{lyk}
) is guaranteed by construction.
  Here I will thus discuss mostly results obtained using the
(symmetric) orbifold methods of ref. \cite{nos}
, although many of the conclusions may be easily  extended to other 4-D string
constructions.

The general idea is the following
\cite{nos}
. One starts with a $(0,2)$  orbifold
compactification of the 10-D heterotic string. It turns out that it is
convenient
to start with the $Spin(32)$ heterotic (instead of $E_8\times E_8$). Models in
which
the gauge group has the structure $G_{GUT}\times G'$, where $G'$ in turn
contains
as a subgroup a copy of $G_{GUT}$ (i.e. $G_{GUT}'\in G'$) are searched for.
We would just have at this point a usual level $k=1$ (0,2) orbifold model with
a
particular
gauge structure. Now we do some kind of modding or projection (to be specified
below)
in such a way that only gauge bosons corresponding to the {\it diagonal}
$G_{GUT}^D$
subgroup of $G_{GUT}\times G_{GUT}'$ survive in the massless spectrum. We are
thus left
with a structure of type $G_{GUT}^D\times G''$ where $G_{GUT}^D$ is realized at
$k=2$.

In fact this scheme is quite general and may be implemented in other classes of
$(0,2)$ 4-D string constructions. For example, it may be used starting with the
class of $(0,2)$ models
\cite{FIQS}
obtained by adding gauge backgrounds and/or
discrete torsion to Gepner and Kazama-Suzuki models
. The final step leading to the
$k=2$ group may be achieved by embedding an order-two symmetry ($\gamma $ in
the
notation of \cite{FIQS} ) by a permutation of $G_{GUT}$ with $G_{GUT}'$.

In the case of orbifolds, three methods in order to do the final step
$G_{GUT}\times G_{GUT}'\rightarrow G_{GUT}^D$ were discussed in refs.
\cite{fiq}
    . In the first
method (I) the underlying $k=1$ model with the $G_{GUT}\times G'$ structure is
obtained by embedding the twist action of the orbifold into the gauge degrees
of
freedom by means of an automorphism of the gauge lattice (instead of a shift).
In $k=1$ models of this type one can have "continuous Wilson-line" backgrounds
\cite{inq2,finq}
which can be added in such a way that the symmetry is broken continuously to
the
diagonal subgroup $G_{GUT}^D$. In the second method (II) one does the final
step
by modding the original model by a $Z_2$ twist under which the two groups
$G_{GUT}$ and $G_{GUT}'\in G'$ are explicitly permuted. The third method (III)
is
field-theoretical. One explicitly breaks the original symmetry down to the
diagonal subgroup by means of an ordinary Higgs mechanism. Although these three
methods look in principle different, there are many  $k=2$ models
wich can be built equivalently from more than one of the above methods.

Giving all the details of these constructions here would be pointless. Let me
just explain
a few features. There are a few of them which are quite general due to their,
in
some
way,  kinematical origin. Consider the mass formula for the left-moving
(bosonic)
string states from any 4-D toroidal orbifold model:
\begin{equation}
{1\over 8}M_L^2\ =\
N_L\ +\ h_{KM}\ +\ E_0\ -1  \ .
\label{ml}
\end{equation}
Here $N_L$ is the left-moving oscillator number,
$h_{KM}$ is the contribution of the KM gauge sector
to the conformal weight of the particle and $E_0$ is the
contribution of the internal (compactified) sector to the
conformal weight.

Let us consider  the case of symmetric
$(0,2)$ Abelian orbifolds. All Abelian $Z_N$ and $Z_N\times Z_M$
orbifolds may be obtained by toroidal compactifications in
which the 6 (left and right) compactified dimensions are
twisted. There are just 13 possible orbifold twists \cite{orb2} which can be
characterized by a shift $v=(v_1,v_2,v_3)$, where $e^{2i\pi v_i}$
are the three twist eigenvalues in a complex basis.
 A consistent symmetric orbifold model is obtained by
combining different twisted sectors in a modular invariant way.
This procedure is well explained in the literature \cite{orb2,imnq} .
To each possible twisted sector there corresponds
a value for $E_0$ given by the general formula:
\begin{equation}
E_0\ =\ \sum _{i=1}^3\ {1\over 2}|v_i|(1-|v_i|)
\label{ecero}
\end{equation}
Notice also that $E_0=0$ for the untwisted sector
which is always part of any orbifold model.
In the case of asymmetric orbifolds
\cite{nsv}
, obtaining $N=1$ unbroken
SUSY allows the freedom of twisting the right-movers
while leaving untouched the (compactified) left-movers.
In this case one can then have $E_0=0$ even in twisted sectors.

Let us go now to the other relevant piece in eq. (\ref{ml}),
namely the contribution $h_{KM}$
of the KM sector to the conformal weight of the particle.
A state in a representation $(R_1,R_2, \cdots)$ will
 have a general weight
\begin{equation}
h_{KM}\ =\ \sum _i {{C(R_i)}\over {k_i+\rho_i}}
\label{peso}
\end{equation}
Here $C(R)$ is the quadratic Casimir of the representation $R$.
$C(R)$ may be computed using $C(R)dim(R)=T(R)dimG$,
where $T(R)$ is the index of $R$. Unless otherwise explicitly stated,
we use the standard normalization in which
$T=1/2$ for the $N$-dimensional representation of $SU(N)$ and
$T=1$ for the vector representation of $SO(2N)$.
With this normalization, for simply-laced groups
the Casimir of the adjoint satisfies $C(A)=\rho.$
The contribution of a $U(1)$ factor to the total
$h_{KM}$ is instead given by $Q^2/k$, where $Q$ is the $U(1)$ charge
of the particle and $k$ is the normalization of the
$U(1)$ generator, abusing a bit it could be called the level
of the $U(1)$ factor.
Formula (\ref{peso}) is very powerful because the
$h_{KM}$ of particles can be computed without
any detailed knowledge of the given 4-D string. This information
is a practical guide in the search for models with
some specific particle content.

Using eq. (\ref{ml}) and the values for $E_0$, $h_{KM}$ computed through eq.
(\ref{ecero}), (\ref{peso})    we can learn,
for instance, what $SU(5)$ or $SO(10)$
representations may appear in the massless spectrum of any possible
twisted  sector of any given Abelian orbifold. In the case of
these  groups we are interested in knowing which twisted
sectors may contain $24$-plets or $45$ and $54$-plets
respectively.

For a $24$-plet
one has $h_{KM}=5/7$; for $SO(10)$
$45$-plets  one has $h_{KM}=4/5$ and, finally,
for  $SO(10)$ $54$-plets  one has
$h_{KM}=1$. From the condition
$h_{KM}+E_0\leq 1$  one draws  the following conclusions:

{\bf i.}
   All those representations  may be present in the untwisted sector
of any orbifold.

{\bf ii.}
  $54$s of $SO(10)$ ($k=2$)  can only be present in
the untwisted sector of symmetric orbifolds.

{\bf iii.}
$45$s of $SO(10)$ ($k=2$) may only appear either in the
untwisted sector or else in twisted sectors of the type
$v=1/4(0,1,1)$ or $v=1/6(0,1,1)$. This is a very restrictive
result since Abelian orbifolds containing these
shifts are limited.

{\bf iv.}
     $24$-plets of $SU(5)$ can never appear in the twisted
sectors of the $Z_3,Z_4,Z_6'$ and $Z_8$ orbifolds.

{}From the above conclusions it transpires that looking
for models with GUT-Higgs fields in the untwisted sector should be
the simplest option, since they can always appear in
$any$ orbifold. This option has another positive aspect in
that the multiplicity of a given representation in the
untwisted sector is never very large, it is always less or equal
than three in practically all orbifolds and is normally equal
to one in the case of $(0,2)$ models. Proliferation of too many
GUT-Higgs multiplets will then be avoided.

To be specific let us present a string $SO(10)$-GUT constructed
from the first of the three orbifold methods mentioned above,
i.e., the "continuous Wilson-line" method. In this method the
orbifold twist $\theta $ is realized in the gauge degrees of freedom in
terms of  automorphisms
$\Theta $.
In the absence of Wilson line backgrounds  $L_i$, the action
of $\Theta$ can be described by an equivalent shift $V$.
In the presence of $L_i$, the embedding is non-Abelian when
$\Theta L_i$ does not give back $L_i$ up to lattice vectors.
When embedding by automorphisms, not all Cartan gauge currents
are given by combinations of derivatives of the 16 bosonic
coordinates $\partial F_I$
since  the lattice coordinates $F_I$ are generically rotated by
$\Theta$   and the unbroken gauge currents must be invariant under
$\Theta$. The Cartan sub-algebra, as well as the step currents, now
arise from $\Theta$ invariant orbits of the $e^{iP\cdot
F}$  operators of the
form
\begin{equation}
 |P\rangle\
+\ |\Theta P\rangle\ +\ \cdots \ +|\Theta ^{N-1}P\rangle
\label{invs}
\end{equation}
 where
$|P\rangle \equiv e^{iP\cdot F}$ and $P^2=2$.
After the continuous Wilson lines are turned on, states not
satisfying $P.L_i=int$ drop out from the spectrum. This
projection kills some Cartan generators thus forcing a reduction
of the rank of the gauge group. This is a necessary condition
to get a residual affine Lie algebra realized at higher level.

Consider \cite{nos} the simplest symmetric orbifold with order 2
symmetries,   namely,
$Z_2\times Z_2$.
 The internal
six-dimensional twists $\theta$ and $\omega$ may be embedded into
the gauge degrees of freedom  by the order two automorphisms $\Theta $ and $
\Omega $
defined by :
\begin{eqnarray}      &
\Theta (F_1,F_2,\cdots ,F_{16})\ =\
(-F_1,-F_2,\cdots ,-F_{10},-F_{11},-F_{12},F_{13},F_{14},F_{15},F_{16}) &
\nonumber \\
  & \Omega
(F_1,F_2,\cdots ,F_{16})\ =\ (-F_1,-F_2,\cdots
,-F_{10},F_{11},F_{12},-F_{13},-F_{14},F_{15},F_{16})&
\label{auti}
\end{eqnarray} The unbroken gauge currents correspond to states $|P\rangle$
with  $P$ invariant plus the oscillators $\partial F_{15},\partial
F_{16}$. Also, from non-invariant $P$'s we can form orbits
invariant  under both $\Theta$ and $\Omega$. Altogether we find 200
currents that can be organized into an $SO(10)\times SO(18)\times
U(1)^2$ algebra realized at level
$k=1$.

Next we turn on a Wilson line background $L$ along, say,
the compactified direction $e_6$. $L$ has the form
\begin{equation}
L\ = \ (\lambda , \lambda  ,\lambda , \cdots ,\lambda ,0,0,0,0,0,0)
\label{wil}
\end{equation}
The parameter $\lambda $ can take any real value
since $L$ is completely rotated by both $\Theta $ and $\Omega $.
The gauge group is broken to $SO(10)\times SO(8)\times U(1)^2$. The
currents
associated to $SO(10)$ are given by
\begin{eqnarray}
	& |{\underline{+1,-1,0,0,\cdots ,0}},0,0,0,0,0,0\rangle\ +\
|{\underline{-1,+1,0,0,\cdots ,0}},0,0,0,0,0,0\rangle &
\nonumber\\ &
\end{eqnarray}
where underlining means that all possible permutations must be
properly considered. One can check that $SO(10)$ is realized at
level $k=2$ whereas $SO(8)$ has $k=1$.

In the untwisted matter sectors $U_1,U_2$ and $U_3$, the corresponding
left-moving vertices transform under $(\Theta , \Omega )$ with
eigenvalues $(-1,1), (1,-1)$ and $(-1,-1)$ respectively. The momenta
involved must also satisfy $P\cdot L = \ int$.
In sectors $U_1$ and $U_2$ there are matter fields transforming as
$(1,8)$ and with different $U(1)$ charges.
In the $U_3$ sector we find the states
\begin{eqnarray}
& \partial F_I\ , \ I=1,\cdots,10  & \nonumber\\
& |\underline{+1,-1,0,\cdots,0},0,0,0,0,0,0\rangle\
-\ |\underline{-1,+1,0,\cdots,0},0,0,0,0,0,0\rangle &
\end{eqnarray}
These states have no $U(1)^2$ charges and belong to a
$(54,1)+(1,1)$ representation of $SO(10)\times SO(8)$.
In $U_3$ we also find
 four extra singlets, charged under the $U(1)$s
only.
Altogether the spectrum of this GUT model is given in
Table \ref{ttres}. The charge $Q$
is non-anomalous whereas $Q_A$ is anomalous.
The gravitational, cubic and mixed gauge anomalies of $Q_A$ are
 in the correct ratios in order to be cancelled
by the 4-D version of the Green-Schwarz mechanism
\cite{gs} .
The degeneracies of the twisted sectors
$\theta ,\omega $ and $\theta \omega $
depend on the way one realizes the
$Z_2\times Z_2$ twist in the compactifying cubic lattice (see ref.
\cite{nos}
  for details).
This model has 4  $SO(10)$ generations and two pairs of $16+{\overline {16}}$
Higgs fields plus additional $10$-plets.

\begin{table}
\begin{center}
\begin{tabular}{|c|c|c|c|c|c|c|}
\hline
$Sector $
& $SO(10)\times SO(8)$ & $Q$ &  $Q_A$ & $A$ & $
B$ &
$C$
 \\
\hline
 $gauginos $ & $(45,1)+(1,28)$ & 0 & 0
& $3m_{3/2}^2$ & $m_{3/2}^2$ & $    0  $
 \\
\hline
$  U_1  $  &     (1,8) &   1/2   &   1/2
& $m_{3/2}^2$ & $m_{3/2}^2$ & $m_{3/2}^2$  \\
\hline
& (1,8)   &    -1/2   &  -1/2
& $m_{3/2}^2$ & $m_{3/2}^2$ & $m_{3/2}^2$ \\
\hline
$  U_2  $ & (1,8)   & -1/2 & 1/2
& $m_{3/2}^2$ & $m_{3/2}^2$ & $m_{3/2}^2$ \\
\hline
& (1,8)    &   1/2  &  -1/2
& $m_{3/2}^2$ & $m_{3/2}^2$ & $m_{3/2}^2$\\
\hline
$   U_3  $ & (54,1)   & 0   &  0
& $m_{3/2}^2$ & $-m_{3/2}^2$ & $-2m_{3/2}^2$\\
\hline
& (1,1)    &    0  & 0
& $m_{3/2}^2$&
$-m_{3/2}^2$ &$-2m_{3/2}^2$\\
\hline
 & (1,1) &  0  & 1     & $m_{3/2}^2$ & $-m_{3/2}^2$ &
$-2m_{3/2}^2$\\
\hline   & (1,1)
&   1  & 0    & $m_{3/2}^2$ & $-m_{3/2}^2$ &
$-2m_{3/2}^2$\\
\hline       & (1,1)
&  -1  & 0    & $m_{3/2}^2$ & $-m_{3/2}^2$ &
$-2m_{3/2}^2$\\
\hline     & (1,1)
&  0  & -1    & $m_{3/2}^2$ & $-m_{3/2}^2$ &
$-2m_{3/2}^2$\\
\hline      $\theta$
& $3(16,1)$   &  1/4  &  1/4   & $m_{3/2}^2$ &   0  &
$-1/2m_{3/2}^2$\\
\hline     &
$(\overline{16},1)$   &  -1/4  &  -1/4   & $m_{3/2}^2$ &    0  &
$-1/2m_{3/2}^2$\\
\hline     $\omega$
& $3(16,1)$   &  -1/4  &  1/4   & $m_{3/2}^2$ & 0    &
$-1/2m_{3/2}^2$\\
\hline         &
$(\overline{16},1)$   &  1/4  &  -1/4   & $m_{3/2}^2$ &   0    &
$-1/2m_{3/2}^2$\\
\hline
$\theta\omega$ &    $ 4(10,1)$   & 0   & 1/2   & $m_{3/2}^2$ & $m_{3/2}^2$ &
$m_{3/2}^2$\\
\hline        &
$4(10,1)$   & 0   & -1/2   & $m_{3/2}^2$ & $m_{3/2}^2$ &
$m_{3/2}^2$\\
\hline       &
$3(1,8)$   & 0   & 1/2   & $m_{3/2}^2$ & $m_{3/2}^2$ &
$m_{3/2}^2$\\
\hline       &
$(1,8)$   & 0   & -1/2   & $m_{3/2}^2$ & $m_{3/2}^2$ &
$m_{3/2}^2$\\
\hline             &
$8(1,1)$   & 1/2   & 0   & $m_{3/2}^2$ & $m_{3/2}^2$ &
$m_{3/2}^2$\\
\hline        &
$8(1,1)$   & -1/2   & 0   & $m_{3/2}^2$ & $m_{3/2}^2$ &
$m_{3/2}^2$\\
\hline
\end{tabular}
\end{center}
\caption{Particle content and charges of the string-GUT example discussed in
the
text. The three rightmost columns desplay three examples of consistent
soft masses from dilaton/moduli SUSY breaking.}

\label{ttres}
\end{table}

There is an interesting feature which turns out to be quite generic
in $SO(10)$ string GUTs obtained from this method (I). In the 0-picture
the full emission vertex operator for the singlet in $U_3$ has the form
\begin{equation}
\partial X_3\ \otimes \ \sum _{I=1}^{10}\partial F_I
\label{svec}
\end{equation}
A Vev for this field precisely corresponds to the Wilson line
background $L$ in eq. (\ref{wil}).
The fact that this background may be varied continuously
means that this singlet is a
{\it string modulus}, a chiral field whose scalar potential is
flat to all orders. Indeed, using the
discrete $Z_2$ R-symmetries of the right-handed sector, it can
be proven that its self-interactions vanish identically.
The GUT Higgs contains the other 9 linear combinations of
$\partial F_I$. These give the diagonal elements
of the symmetric traceless matrix chosen to represent the 54-plet.
the associated vertex operator is
\begin{equation}
\partial X_3\ \otimes \ \sum _{I=1}^{10} c_I\partial F_I
\ \ ;\ \ c_I\in {\bf R}, \  \ \sum_Ic_I=0 \ .
\label{hvec}
\end{equation}
Vevs for these nine components of the 54 would correspond to the
presence of more general Wilson backgrounds of the form
$L=(\lambda_1,\lambda_2,\cdots ,\lambda_{10},0,0,0,0,0,0)$ with
$\sum _{I=1}^{10} \lambda_I=0$.
These more general backgounds break the symmetry further to some
$SO(10)$ subgroup like $SU(4)\times SU(2)_L\times SU(2)_R$.
The fact that these other nine modes may
be continuously varied means that they are also string
moduli or, more generally, that the 54-plet of $SO(10)$ in this
model is itself a string modulus! We find that this property of the
GUT-Higgs behaving as a string modulus, on equal footing
with the compactifying moduli $T_i$, is very remarkable.

This example  belongs to a whole
class of models obtained through continuous Wilson lines.
A general characteristic is that they are $SO(10)$ models
in which the GUT Higgs is a 54 multiplet.
Moreover, there is only one such GUT Higgs coming from
the untwisted sector and behaving like a string modulus.
On the other hand, the rest of the particle content is model
dependent. This includes the number of generations,
existence of Higgses 10s, $(16+{\overline {16}})$s,
hidden gauge group, etc. For instance, the number of generations
can be changed by adding discrete Wilson lines to the original
orbifold.

The second orbifold method (II), which is implemented by permutations, is more
versatile
\cite{prep}
{}.
In this case one may obtain models similar to the previous one both with either
one
$54$ or one $45$ in the untwisted sector. One can also find $SU(5)$ models with
adjoints $24$s  in the untwisted sector (sometimes also in some  twisted
sectors).
Instead of showing more examples it is perhaps more interesting to desplay some
general properties and selection rules
\cite{prep}
which one can derive for this kind
of string-GUTs. As  will be clear, some of those will be more general and apply
to
any string-GUT constructed through any method.
\bigskip

{\it General selection rules for any k=2 string-GUT}
\bigskip

{\bf i)}  All superpotential terms have $dim\geq 4$ (i.e., no mass terms).

{\bf ii)} At $k=2$ the only reps. which may be present in the massless spectrum
are:
$5,10,15$ and $24$-plets for $SU(5)$; $10,16,45$ and $54$-plets for $SO(10)$.

{\bf iii)} $SO(10)$: The rep. $\underline {54}$ for $k=2$ is special. All its
left-handed conformal weight comes from the KM sector. As a consequence:
a) A $\underline {54}$ cannot be charged under any $U(1)$ nor any other gauge
group.
b) Couplings of the type $X({\underline {54}})({\underline {54}})$ or
$X({\underline {54}})({\underline {54}}')$ (where $X$ is some singlet) are not
allowed.

In addition to these general rules one can prove several other ones for
string-GUTs
obtained from symmetric orbifolds. Some of these are as follows:
\bigskip

{\it Selection rules for  k=2 string-GUTs from symmetric orbifolds}
\bigskip

{\bf i)} There are no selfcouplings ${\underline {54}}^n$ for any n.

{\bf ii)} There cannot be couplings of type $({\underline {54}})({\underline
{45}})
({\underline {45}})$.

{\bf iii)} There cannot be $SU(5)$ self-couplings of type ${\underline
{24}}^3$.

In practice, when constructing explicit models in symmetric orbifolds, the
constraints are
even  tighter. As we mentioned above, there is normally just one GUT-Higgs
in $SO(10)$, either a $\underline {54}$ or a $\underline {45}$ in the untwisted
sector.
Being in the untwisted sectors, selfcouplings of the type ${\underline {45}}^n$
or  couplings of type $X({\underline {45}})({\underline {45}})$ are
also forbidden (see ref.
\cite{prep}
 for a more detailed explanation of selection rules).
Many of these couplings have been used in the past in SUSY-GUT model
building in order to trigger GUT-symmetry breaking while obtaining as the
low energy sector the MSSM. With the above type of constraints it seems it is
very difficult (if not impossible) to construct models whose low energy sector
is
indeed the MSSM. The absence of some relevant GUT-Higgs selfcouplings cause
extra chiral multiplets to remain massless. That will be the case of the
GUT-partners of the Goldstone bosons of GUT-symmetry breaking. For example,
upon symmetry breaking by an adjoint $\underline {24}$, twelve out of the
24 fields remain massless. They transform as
\begin{equation}
  (8,1,0)\ +\ (1,3,0)\ +\ (1,1,0)
\label{octete}
\end{equation}
under $SU(3)\times SU(2)\times U(1) $. This seems quite a generic situation
which I would expect to be present in more general 4-D string constructions
like asymmetric orbifolds or models based on the fermionic construction.
In these more general cases some of the above strict selection rules are
relaxed in principle but not very much in practice. For example, if the
$k=2$ model comes from the diagonal sum of two $SO(10)$ $k=1$ factors, the
$45$s or $54$s obtained originate from a $(10,10)$ rep. of the original theory.
Such reps. do not admit cubic selfcouplings  and the same is
expected for the $54$ (due to its antisymmetry there are no cubic couplings for
the
$45$ anyhow). Since the extra particles above will have masses only of the
order
of
the weak scale,  they will sizably contribute to the running of the gauge
coupling
constants.
One can  perform a one-loop analysis of the running of the
gauge coupling constants and check  that, with the
particle content of the minimal SUSY-SM plus the additional
fields above, there is no
appropriate gauge coupling unification
in the vicinity of $10^{15}-10^{17}$ GeV. In the case of
$SO(10)$ an intermediate scale of symmetry breaking could
improve the results. We thus see that gauge coupling
unification is not particularly better in string-GUTs
than in direct SM string unification, if the above
analysis is correct.

The most severe problem of SUSY-GUTs is the infamous
doublet-triplet splitting problem of finding a mechanism to
understand why, for example, in the $5$-plet Higgs of $SU(5)$
the Weinberg-Salam doublets remain light while their
coloured triplet partners become heavy enough
to avoid fast proton decay.
 The most simple,
but clearly unacceptable, way to achieve the splitting is
to write a term in the $SU(5)$ superpotential
\begin{equation}
W_H\ =\ \lambda
	H\Phi _{24} {\bar H}\ +\ MH{\bar H}
\label{sup5}
\end{equation}
and fine-tune $\lambda $ and $M$ so that the doublets turn
light and the triplets heavy. Since there are no
explicit mass terms in string theory this inelegant
possibility is not even present. Another alternative
suggested long time ago is
the ``missing partner" mechanism
\cite{miss} . Formulated in
$SU(5)$ it requires the presence of $50$-plets
in the massless sector which is only possible
for level $k \geq 5$, a very unlikely possibility
\cite{fiq,nanop} .
A third mechanism, put forward in the early days of SUSY-GUTs,
is the ``sliding singlet" mechanism
\cite{slid,iruno} . This requires the existence of
a singlet field $X$, with no self-interactions, entering in the
mass term in eq. (\ref{sup5}). $W_H$ is then replaced by
\begin{equation}
W_X\ =\ \lambda
	H\Phi _{24} {\bar H}\ +\ XH{\bar H}   \ .
\label{supx}
\end{equation}
The idea is that the vev of the $24$ is fixed by
other pieces in the potential
 but the vev
of $X$ is undetermined to start with, i.e. the vev ``slides".
Now, once the electroweak symmetry is broken by the
vevs of $H,{\bar H}$, the minimization conditions give
$\lambda \langle -{3\over 2}v\rangle+\langle X\rangle=0$
where diag($\langle \Phi _{24}\rangle)
=v(1,1,1,-3/2,-3/2)$. In this way
$X$ precisely acquires the vev needed for massless
doublets. This is in principle a nice dynamical
mechanism but it was soon realized that it is
easily spoiled by quantum corrections
\cite{slidprob,nil} .

Interestingly enough, one finds  that
in string GUTs, couplings of the ``sliding singlet" type are frequent,
the main difference now being that the GUT-Higgs field also ``slides".
In particular, this happens in models in which
the GUT-Higgs is a modulus, as in the examples discussed
above.
Take for example the
$SO(10)$ model discussed above whose massless spectrum is
displayed in the table . The singlets in the $U_3$ sector $S^0=(1,1)_{0,0}$,
$S^+=(1,1)_{0,1}$, $S^-=(1,1)_{0,-1}$ do also behave as moduli. Both
these singlets and the $54$ couple to the decuplets
$H^+=(10,1)_{0,1}$ and $H^-=(10,1)_{0,-1}$. The sub-indices in all
these fields refer to their $Q$ and $Q_A$ charges. It is easy to check
that there are flat directions in this scalar moduli space
in which the gauge symmetry is broken down to
$SU(4)\times SU(2)\times SU(2)$ and some of the doublets remain light
whereas the colour triplets remain heavy (the symmetry is broken
down to the SM through the vevs of the $16+{\bar {16}}$ pairs).
If the sliding-singlet argument were stable under quantum corrections,
the regions in moduli-space in which there are light doublets
would be energetically favoured.

As the above example shows, the appropriate language to describe
the doublet-triplet splitting problem within the context of the
above string-GUTs is in terms of the scalar moduli space of the
model. At generic points in the moduli space there are no massless
Higgs doublets at all, they are all massive. At some
``multicritical" points in moduli space some Higgs fields
become massless. This is very reminiscent of the behaviour of
the moduli spaces of other well studied string moduli, those
associated to the size and shape of the compact manifold usually
denoted by $T_i$. It is well known that generically there are
points in the $T_i$ moduli space in which extra massless fields
appear. This is also apparently the case of the
moduli space associated to the dilaton complex field $S$. The
problem of understanding the doublet-triplet splitting within this
context would be equivalent to finding out why we are sitting on a
region of moduli space in which massless doublets are obtained.
It could well be that an appropriately modified version of the
sliding-singlet mechanism is at work and that region of moduli
space is energetically favoured.

 To summarize this section,
 I  believe that the doublet-triplet splitting problem
is a crucial issue and should be addressed in any model before trying
to extract any further phenomenological consequences such as
fermion masses. It is also important to understand whether it
is possible to build string GUTs in which the massless sector
is just the MSSM, or else whether
the existence of extra massless chiral fields is really generic.
This would dictate the necessity of intermediate scales to
attain coupling constant unification.

\section{Soft SUSY-breaking terms from dilaton /moduli sectors}

Let us turn now to a different subject. The idea is trying to extract
some information about the structure of effective SUSY-breaking soft terms
which are left out once supersymmetry is spontaneously broken.
This is a very important objective since by the year 2005 the
spectrum of supersymmetric particles might be tested at LHC
and this could  be one of the few experimental windows we could have
to test the theory. Since very
little is known about the origin of supersymmetry-breaking in string theory,
this aim looks impossible.
This is not as hopeless as it seems. We do not need to know {\it  all}
the details of a symmetry-breaking process in order to get important
physical information. A well known example of this is the standard model
itself.
We do not know yet how $SU(2)\times U(1)$ breaking takes place. But, assuming
that somehow a (composite or elementary) operator with the quantum numbers of
a doublet gets a vev, we get a lot of information.

The idea is to apply a similar philosophy for SUSY-breaking in string theory
\cite{CFILQ,il}
{}.
We have to try and identify possible chiral fields $\phi _i$ such that their
auxiliary fields $F_i$ could get non-vanishing vev and break SUSY. In string
models
there are some natural candidates to do the job: the complex dilaton
$S={{4\pi }/{g^2}+i\theta }$ and the moduli fields $T_i$ whose vevs determine
the size and
shape of the compact space. The field $S$ is present in any 4-D strings and the
$T_i$ fields at least in any model obtained from compactification. Thus these
singlet fields are generic in large classes of string models. An additional
advantage
is that these fields couple to matter only with non-renormalizable couplings
supressed by powers of $1/M_{Planck}$. This is a condition which is required
in supergravity models with supersymmetry breaking in a" hidden sector".
The scalar potential for $S$ and $T_i$ is flat order by order in perturbation
theory and it is expected that non-perturbative effects will i) induce
a non-trivial scalar potential for those fields yielding
$<S>\not= 0$, $<T_i>\not= 0$ and ii) break supersymmetry spontaneously.

The crucial assumption here is to locate the origin of SUSY-breaking in the
dilaton/moduli sector
\cite{CFILQ,il,KL,BIM}
. It is perfectly conceivable that other fields in the
theory, like charged matter fields, could contribute in a leading manner to
supersymmetry
breaking. If that is the case the structure of soft SUSY-breaking terms will be
totally
model-dependent and we would be able to make no model-independent statements at
all
about soft terms. On the contrary, assuming the seed of SUSY-breaking
originates
in the dilaton-moduli sectors will enable us to make some predictions which
might be
testable. We will thus make that assumption without further justification.
Let us take the following parametrization
\cite{BIM,BIMS}
for the dilaton/moduli auxiliary
fields $F^S$ and $F^i$ vevs:
\begin{equation}
{{G^S_S}^{1/2} {F^S}}\ =\ \sqrt{3}m_{3/2}sin\theta \ \ ;\ \
{{G^i_i}^{1/2}{F^i}}\ =\ \sqrt{3}m_{3/2}cos\theta\ \Theta _i \ \
\label{auxil}
\end{equation}
where $\sum _i \Theta _i^2=1$ and $m_{3/2}$ is the gravitino mass.
The angle $\theta $ and the $\Theta _i$ just parametrize the
direction of the goldstino in the $S,T_i$ field space.
This parametrization has the virtue that when we plugg it in the
general form of the supergravity scalar potential, its vev (the cosmological
constant ) vanishes by construction. We now need some information about the
couplings of the dilaton/moduli fields. Those are given by the Kahler potential
$G$ and the gauge kinetic functions $f^a$. The latter are given at the tree
level
by $f^a=k^aS$, where $k^a$ are the Kac-Moody levels, for any 4-D string. The
tree-level kahler
potential for $S$ is $-log(S+S^*)$, also for any model. The kahler potential
for the moduli are more complicated and model-dependent and we need to specify
the class
of models and moduli that we are considering. We will concentrate here
on the three untwisted moduli $T_1,T_2,T_3$ always present in
$(0,2)$ toroidal symmetric orbifold constructions (for particular examples
there may be additional untwisted moduli  and  also complex
structure structure  fields $U_1,U_2,U_3$).
 For this large class of
models one can
write for the kahler potential \cite{WIT,DKL}
\begin{equation}
K(S, T_i,C_r..)\ =\ -log(S+S^*)\ -\ \sum _i log(T_i+T_i^*)\ +\
\sum _n |C_r|^2 \Pi_i(T_i+T_i^*)^{n_r^i}
\label{kahler}
\end{equation}
where $C_r$ are charged chiral fields and  the $n_r^i$ are fractional numbers
called "modular weights" which depend on the given field $C_r$. The sum in
$i$ may be extended to all the three moduli $T$-fields and also to the
complex-structure
$U$-fields. Plugging this information into the supergravity lagrangian one
finds
the following results for the scalar masses, gaugino masses and soft trilinear
coupling $A_{rst}$ associated to a Yukawa coupling $Y_{rst}$ \cite{BIM,BIMS} :
\begin{eqnarray}
 & m_r^2 =\  m_{3/2}^2(1\ +\ 3cos^2\theta\ {\vec {n_r}}.{\vec {\Theta ^2}})   &
\nonumber\\
 & M^a =\  \sqrt{3}m_{3/2}sin\theta  &
\nonumber\\
&  A_{rst} =\   -\sqrt{3} m_{3/2}[sin\theta \ +\ cos\theta ({\vec {\Theta
}}.({\vec u}+
{\vec {n_r}}+{\vec {n_s}}+{\vec {n_t}}))] \ . &
\label{softy}
\end{eqnarray}
Here we have used vectorial notation in the space of the $T-U$ moduli and
we define ${\vec u}=(1,1,...)$. (An additional term should be added to
$A_{rst}$ for Yukawa couplings $Y_{rst}$ depending on the moduli).
Several observations are in order. Fist of all, in the case of dilaton
dominance in the SUSY-breaking process ($sin\theta =1$) one gets
the remarkably simple universal\cite{il}  result \cite{KL,BIM} :
\begin{eqnarray}
& -A_{rst}  =  \ M^a=\sqrt{3} m_{3/2}  & \\
&  m_r^2   =  \  m_{3/2}^2  . &
\label{dilat}
\end{eqnarray}
This result in fact applies for any 4-D string (not only orbifolds)
whenever the dilaton dominates. A second observation is that a similar
structure
of soft terms is obtained  for orbifold models in which
${\vec n_r}.{\vec {\Theta ^2}}=-1/3$ and ${\vec {n_r}}+{\vec {n_s}}+{\vec
{n_t}}
=-{\vec u}$. This happens for any value of $\theta $ (i.e., not
necessarily dilaton dominance)
if one assumes $F^1=F^2=F^3$ (equal SUSY-breaking contribution from the three
untwisted $T$-fields, ${\vec {\Theta ^2}}=(1/3,1/3,1/3)$
) and that the charged fields $C_r$ have overall modular weight
$n_r={\vec u}.{{\vec n_r}}=-1$
(this happens for all particles in the untwisted sector and  particles in
some types of twisted sectors).
A third important observation is that
the $mass^2$  of the scalar fields is not positive definite and hence
there  are choices of $sin\theta $ and the ${{\vec {\Theta ^2}}}$ for which
tachyons may appear \cite{CFILQ,BIM}  .

A simplified case in which only the
dilaton $S$ and the "overall modulus" $T$ contribute to SUSY-breaking
(i.e.${{\vec {\Theta ^2}}}=(1/3,1/3,1/3)$)
was studied in ref.\cite{BIM} . In this case one finds that 1) the particles
in both untwisted and twisted sectors with overall modular weights $n=-1$
never become tachyonic; 2) In order to avoid particles with smaller
modular weights (i.e. $n=-2,-3..$) to become tachyionic one has to confine
oneself to goldstino angles $cos^2\theta \leq 1/|n| $; 3) Due to these
constraints one always has bigger gaugino than scalar masses ($M^2\geq m_i^2$).
There is only one situation in which the gauginos may become lighter than the
scalars. This happens
\cite{BIM}
when the chiral fields have all modular weights $n=-1$
and $sin\theta \rightarrow 0$, in which case both $M,m\rightarrow 0$ and
including loop corrections to $G$ and $f^a$ can reverse the situation and yield
gaugino masses  smaller than scalar masses.

Considering the more general case
\cite{BIMS,JAPSD}
in eq. with several moduli $T_i$
modifies somewhat the general conclusions. One finds that 1) particles with
overall modular weight $n=-1$ can also become tachyonic for some choices
of the angles and  2) The gaugino masses may become lighter than scalar
masses even at the tree level. As an example of the possibilities offered
let me consider some consistent choices of soft mass terms for
the string $SO(10)$ GUT discussed above. For simplicity
I only consider the possibility of the $S$ and the $T_i$ $i=1,2,3$ , (and not
the
$U_i$) contributing to SUSY-breaking. Since it is a $Z_2\times Z_2$
orbifold, the twisted modular weights are ${\vec n}_T=(
{\underline{{{-1}\over 2},{{-1}\over 2},0}})$, where the underlyning means
permutations. The untwisted modular weights are as usual
${\vec n}_U=({\underline {-1,0,0}})$. The three rightmost columns in the table
show three consistent choices of soft masses: A) Dilaton dominated case
($sin\theta =1 $). All the scalars have the same mass, $\sqrt{3} $ times
lighter
than the gauginos. B) Case with $sin^2\theta =1/3$ and
${{\vec {\Theta ^2}}}=(0,0,1)$. In this case the gauginos and Higgs $10$-plets
have
equal masses,the $16$-plets have zero mass and the GUT-Higgs have negative
$mass^2$.
The latter property is interesting since it show us that SUSY-breaking may
authomatically trigger GUT-symmetry breaking; C) Case with $sin\theta =0$
and ${{\vec {\Theta ^2}}}=(0,0,1)$. In this case the dilaton does not
contribute
to SUSY-breaking and the gauginos are massless at the tree-level. The Higgs
$10$-plets have positive $mass^2$ but the GUT-Higgs and the $16$-plets
get negative $mass^2$. The latter may also enforce that there are
$16+{\bar {16}}$ pairs getting a vev. Thus we see that a variety of
phenomenological
possibilities open up depending on what is the role of the different moduli
in the process of SUSY-breaking.

It must be emphasized that, given a specific
string model, there is only certain type of soft terms which can be added
consistently
with the assumptions of dilaton/moduli dominance in SUSY-breaking. Taking
a random choice of soft terms would lead to inconsistencies. For example there
is
always a rule
\cite{BIMS}
which connects the soft masses of particles in the three
untwisted sectors $i=1,2,3$ with that of the gauginos,
$m_1^2+m_2^2+m_3^2=M^2$ (see the discussion below).
 The reader may check this constraint in the three examples
in the table.
In fact a similar sum-rule is still correct for twisted particles with
overall modular weight $n=-1$, and not only for untwisted fields.
More details and examples can be found in a forthcoming publication
\cite{BIMS}  .

\section{Dilaton-induced SUSY-breaking is special}

Indeed the soft terms relationships obtained under the assumption of
dilaton- dominance SUSY-breaking is special in several respects. First,
these boundary conditions for soft terms are obtained for {\it any}
4-D $N=1$ string and not only for orbifolds. Secondly, these
conditions are universal, gauge group independent and flavour independent.
Thirdly, the soft masses obtained for scalars are positive definite, lead to
no tachyons. This is to be compared with situations in which other fields like
the moduli contribute to SUSY-breaking. We have seen how easy is to get
negative $mass^2$ when the moduli contribute to SUSY breaking
\cite{CFILQ,BIM}
. We all hope
that, whatever the string theory describing the spontaneously broken
SUSY phase could be, it will be a tachyon-free modular invariant theory.
Dilaton-dominance SUSY-breaking is not necessary for the absence of
tachyons, there are also situations in which e.g., the moduli dominate
and still there are no tachyons.  But dilaton dominance guarantees the
absence of tachyons. So also in this sense dilaton induced SUSY-breaking
is special.

Dilaton SUSY-breaking is special in yet another aspect, which has past
mostly unnoticed in the literature. It has been recently realized
\cite{IM,JJJ}   that the
boundary conditions $-A=M=\sqrt{3}m$ of dilaton dominance coincide with
some boundary conditions considered by Jones, Mezincescu and Yao in 1984
\cite{JMY}
in a completely different context. It is well known that one
can obtain two-loop finite $N=1$ field theories by considering appropriate
combinations of matter fields (so that the one-loop $\beta $-function
vanishes) and Yukawa couplings (so that the matter field anomalous dimensions
vanish). It has also been argued in favour of  the complete finiteness of this
type
of theories to all orders.
What Jones, Mezincescu and Yau did is to
look for SUSY-breaking soft terms which do not spoil one-loop finiteness when
added to these finite theories. They came out with universal soft terms with
$-A=M=\sqrt{3} m$. It was also shown in 1994 by Jack and Jones that
two-loop finiteness was also preserved by this choice \cite{JJM} .

This coincidence is at first sight quite surprising since
we did not  bother
about the loop corrections when extracting these boundary conditions from
the dilaton dominance assumption. Also, effective $N=1$ field theories from
strings do not in general fulfill the finiteness requirements (in fact I do not
know of any which does). Why dilaton-dominance bothers to yield
soft terms with such improved ultraviolet behaviour?

A heuristic motivation goes as follows.
The dilaton sector in a 4-D string is completely model independent (at
least at the tree-level). Hence the gauge kinetic function $f_a=k_aS$ and
$G(S,S^*)=-log (S+S^*)$ for {\it any} 4-D string, independently of e.g.,
what compactification we used to obtain it. This means that, if the
assumption of dilaton-dominance makes sense at all, it has to lead to soft
terms which are consistent with {\it any } possible compactification and
also has to be independent of the particular choice of compactification.
In particular, the obtained soft terms have to be consistent with the
simplest of all kinds of compactifications, a toroidal compactification
preserving $N=4$ supersymmetry. What do I mean by soft terms consistent
with $N=4$ supersymmetry? By that I mean that the soft terms should be in
the list of terms which mantain the finiteness 	 properties of $N=4$
supersymmetry. The reason for that is that, if there is indeed some
mechanism by which SUSY is {\it spontaneously} broken in the dilaton
sector one does not expect  the induced soft terms below the SUSY-breaking
scale  to produce  new logarithmic divergences in a theory ($N=4$ SUSY) which
was originally finite.

The types of SUSY-breaking soft terms which may be added to
$N=4$ SUSY without spoiling finiteness is well known
\cite{soft4}
. First, one can add $N=1$ preserving
masses for the $N=1$ chiral multiplets contained in $N=4$. These are not very
interesting
since we are interested in soft terms leaving no unbroken supersymmetry.
A second more interesting possibility is to add soft masses $M$ for the
gauginos along with masses  $m_1^2,m_2^2,m_3^2$ for the three multiplets
of adjoint scalars present in the theory verifying:
\begin{equation}
m_1^2\ +\ m_2^2\ +\ m_3^2\ =\ M^2 \ .
\label{N=4uno}
\end{equation}
This constraint may be interpreted just as $Supertrace(Mass)^2=0$
for$N=4$. In addition, the presence of gaugino masses generates logarithmic
divergences involving a holomorphic trilinear coupling (an $A$-term) which
is only cancelled if \begin{equation} M\ =\ -A \ . \label{AN=4}
\end{equation} I already mentioned how a sum-rule like (\ref{N=4uno} )  is
indeed verified  by the dilaton/moduli induced soft terms involving the
{\it untwisted } sector particles in orbifolds. In the case of a
model-independent source of SUSY-breaking like the one we are discussing
one expects $m_1^2=m_2^2=m_3^2=m^2$. So, one thus arrives for consistency
with $N=4$ at the universal conditions $-A=M=\sqrt{3}m$.  One thus can
understand the dilaton-induced boundary conditions as a consistency
condition due to the fact that i) dilaton couplings are
compactification-independent and  thus 2)  should obey consistency
constraints from the most constrained  compactifications, $N=4$ preserving
compactifications.  Notice that in an $N=1$ theory the sum rule
(\ref{N=4uno} )
 will not be  in general preserved,
it is the boundary conditions $-A=M=\sqrt{3}m$ which generalize to the	$N=1$
case,
not the $N=4$ expressions themselves.

Coming back to the finiteness properties of this type of soft terms, it is
clear that if we had an $N=1$ two-loop finite theory as the effective low
energy
theory from some 4-D string model, dilaton SUSY-breaking would respect these
finiteness properties. On the other hand there is no reason for an $N=1$
theory from strings to be finite as a field theory, it is already finite
anyhow due to the string cut-off (modular invariance). If we add these soft
terms to a {\it non-finite} $N=1$ theory the ultraviolet properties do not
specially improve, but at the scale at which those relationships hold
($M_{string}$) one finds e.g. that the $\beta $ functions associated to the
soft terms are proportional to the $\beta $-function of the Yukawa
couplings h, i.e., \begin{eqnarray}
\beta _A(h,g^2,M,m,A)|_{M_{string}}\ &\propto \ M {\beta }_h(h,g^2) & \nonumber
\\
\beta _{m^2}(h,g^2,M,m,A)|_{M_{string}}\ &\propto \ M^2/h {\beta }_h(h,g^2)  &
\label{betas}
\end{eqnarray}
So the theory becomes finite if the underlying unbroken-SUSY theory was finite.
On the other hand, if we start from a finite $N=1$ theory and we add
random soft terms one finds that ${\beta }_A\propto h^2(A+M)$
and $\beta _{m^2}\propto h^2(3m^2-M^2) $ so that the dilaton-dominated
boundary conditions would constitute a fixed point of the renormalization
group equations. This again shows us the special properties of this
choice of soft terms.

The above discussion shows that the assumption that the auxiliary field
associated to the dilaton breaks supersymmetry leads to soft terms with
remarkable finiteness properties. This fact seems to be related to the
$S-duality$ structure of the dilaton couplings. For example,
one can check that the
result $M=-A$ is obtained in the dilaton dominated scheme due to the fact that
the
following functional expression is verified :
\begin{equation}
f(S)^S\ =\  2Ref(S){G(S,S^*)_S^S}^{1/2}
\label{cur}
\end{equation}
where $f$ and $G$ are the gauge kinetic function and $S$-field Kahler
potential.
This relationship is related to the $SU(1,1)$ structure of the field $S$ in the
$N=4$
supergravity Lagrangian
\cite{dero}
. So the dilaton-dominated soft terms are intimately
connected to the $S-duality$ symmetry \cite{sdual,sen}
 underlying these theories.

Recently Seiberg and others
\cite{seiberg}
have discussed the existence of certain duality properties
between different classes of $N=1$ theories with different particle content
and with or without Yukawa couplings.
It would be desirable to see to what extent their results could be extended to
the
$N=0$ case. An obvious first step in that direction would be the addition of
SUSY breaking soft terms.
In view of the above discussion it is reasonable to think
that soft terms of the type $-A=M=\sqrt{3}m$ could have an special status in
this
respect.
Notice also that the proportionality between soft terms and Yukawa
$\beta $-functions at $M_{string}$ shown in
eq.(\ref{betas}) shows
the presence of a non supersymmetric marginal operator structure
analogous to the $N=1$ examples discussed in
ref. \cite{leigh}  .

\section{Oulook and speculations}

The above lines discussed several directions recently explored in trying to
establish contact between the physics at the string scale and the physics
at the weak scale, which is the one amenable to experimental test. It is
important to realize that by the year 2005 the LHC should provide us with
important experimental information about the origin of the weak scale.
If low energy supersymmetry is correct, the spectrum of SUSY particles
should be tested. We do not have at the moment a theory of supersymmetry
breaking but we still have ten years ahead to find one! I certainly believe
that it should be easier to find a theory of soft terms rather than a theory
of fermion masses. At least, it seems that the former could have a more
model-independent origin than the second.

In the previous lines I parametrized SUSY-breaking in terms of the vacuum
expectation
values of the auxiliary fields of the dilaton and moduli but I never discussed
what
could be the dynamical origin of supersymmetry breaking. I did not discussed
either
what is the dynamics which fixes the vevs $<S>$ and $<T_i>$. The most popular
scenarios assume that the same dynamics which break SUSY at the same time
fix those vevs. It is not clear to me that this is necesarily the case. It is
conceivable that some string dynamics could fix $<S>$ and/or $<T_i>$ to be
of order the string scale and then some low energy field-theoretical effect
(e.g., gaugino condensation) could break supersymmetry \cite{IQ} . It is not
clear
what string effects could fix the $S,T$ vevs without breaking supersymmetry
at the string scale, but one can use the duality symmetries associated to those
fields to restrict the possibilities. Indeed, the well known $T$-duality
symmetries
would suggest that the most natural values for $<T>$ should be around the
selfdual point, $<T>\simeq 1$ and that kind of result is obtained in
$T-duality$-invariant versions of gaugino condensation
\cite{sem}
. If some sort
of $S-duality$
\cite{sdual}
is correct in $N=1$ theories, one should also expect
$<S>\simeq 1$. I would argue that this is not necesarily unreasonable if the
massless sector of the theory contains particles beyond the ones in the MSSM,
which is in fact the generic case in explicit string models. In this
case the gauge couplings will not be asymptotically free and may become
quite large at the string scale.

Things may be more complicated than the tacit assumption hidden in the
previous sections, that we are in a perturbative regime of the string.
 It could well be that the non-perturbative string effects modify in a
substantial manner all the perturbative 4-D backgrounds that we are using
at the moment in explicit constructions. In this case one can still hope
that these corrections do not modify substantially the $N=1$ superpotentials
but only the D-terms (see talk by M. Dine in these proceedings).
If we are less lucky even the superpotentials could be affected. This would
make rather difficult to extract predictions from any 4-D string construction
unless we know all the relevant non-perturbative dynamics, something which
looks
rather remote. On the other hand this possibility would have in my opinion
one interesting aspect (probably the only one!). Standard 4-D strings
are always excesively rigid in providing Yukawa couplings. They tend to have
so many continuous and discrete symmetries that many (too many) couplings
(including non-renormalizable ones) are forbidden. A typical example of this
is the absence of selfcouplings of the GUT-Higgs discussed above.
Perhaps string non-perturbative effects could generate new superpotential
terms (e.g., like ${\underline {24}}^3$ in $SU(5)$)
which could be absent in the perturbative vacuum one started from.

The other tacit assumption is that we are identifying correctly the
short distance
elementary degrees of freedom of the standard model in trying to embed it into
a 4-D string. The recent results  by
Seiberg and others
\cite{seiberg}
show how two  different $N=1$ theories with different gauge
group could be dual to each other and describe the same physics in the
infrared.
An example of this is the equivalence of the physics of a $N=1$ $SU(N)$
theory with $N_f$ flavours at weak coupling to the physics of an $SU(N_f-N)$
also with
$N_f$ flavours at strong coupling. Thus , as suggested in
the first article in \cite{seiberg}
, perhaps all or part
of the known elementary particles of the standard model are in fact dual to the
truly elementary particles at short distances. This is a very intriguing
possibility.
It could well be that e.g., the $SU(3)$ colour interactions and the quarks
were not elementary but dual to the true elementary states. It would be the
latter
states which we should unify along with the $SU(2)\times U(1)$
interactions and the leptons into a string theory. The whole hypothesis
of the "desert'' should then be reconsidered, including its emblematic
prediction,
gauge coupling unification, since it would not be the observed $\alpha _s$
coupling which should unify with the other two, but the dual.  Although one
cannot
directly apply the arguments of ref.
(\cite{seiberg})
 to a non-semisimple chiral theory
like the SM, it is amussing to note that for SUSY-QCD with the observed 6
flavours
$N=N_f-N=3$ and hence the gauge group would be the same in the dual theory.



\end{document}